\documentclass[twocolumn,amstext,amsmath,amssymb,prl,preprintnumbers,superscriptaddress,floatfix,showpacs,nofootinbib]{revtex4}

\usepackage[dvips]{epsfig}
\usepackage{slashed}
\usepackage{mathrsfs}
\usepackage{graphicx}
\usepackage{xcolor}
\usepackage{hyperref}
\usepackage{dcolumn}
\usepackage{subfigure}
\usepackage{xspace}
\usepackage{ulem}

\begin{document}
\preprint{LMU-ASC 18/14}
\preprint{TUM-HEP 936/14}
\preprint{FLAVOUR(267104)-ERC-71}

\title{Dark Matter Stability without New Symmetries}

\author{Oscar Cat\`a}
 \affiliation{Ludwig-Maximilians-Universit\"at M\"unchen, 
   Fakult\"at f\"ur Physik,\\
   Arnold Sommerfeld Center for Theoretical Physics, 
   80333 M\"unchen, Germany}
  \affiliation{TUM-IAS, Lichtenbergstr. 2a, D--85748 Garching, Germany}
\affiliation{Physik-Department, Technische Universit\"at M\"unchen, James-Franck-Stra\ss{}e, 85748 Garching, Germany}

  \author{Alejandro Ibarra}
\affiliation{Physik-Department, Technische Universit\"at M\"unchen, James-Franck-Stra\ss{}e, 85748 Garching, Germany}


\begin{abstract}
The stability of dark matter is normally achieved by imposing extra symmetries beyond those of the Standard Model of Particle Physics. In this paper we present a framework where the dark matter stability emerges as a consequence of the Standard Model symmetries. The dark matter particle is an antisymmetric tensor field (analogous to the one used for spin-1 mesons in QCD), singlet under the Standard Model gauge group. The Lagrangian possesses an accidental $Z_2$ symmetry which makes the dark matter stable on cosmological time scales. Interactions with the Standard Model fields proceed through the Higgs portal, which allows the observed dark matter abundance to be generated via thermal freeze-out. We also discuss the prospects for observing this dark matter particle in direct detection experiments.
\end{abstract}

\maketitle

Among the myriad of currently identified matter particles, only very few have cosmologically long lifetimes: the proton, the electron and the three neutrinos. The longevity of these particles is well understood in the Standard Model framework and arises as a natural consequence of the Lorentz and gauge symmetries of the model: the electron and the lightest neutrino are predicted to be absolutely stable, due to electric charge and angular momentum conservation, respectively, while the proton and the two heavier neutrinos can only decay into lighter particles when higher-dimensional operators (presumably generated by new physics at a high-energy scale) are added to the renormalizable Lagrangian.

On the other hand, there is mounting astrophysical and cosmological evidence for the existence of (at least) one more cosmologically-stable particle not included in the Standard Model, the dark matter particle~\cite{Bertone:2004pz}. Light dark matter candidates, such as the axion or the sterile neutrino, are naturally long-lived due to the suppression of the decay rate by the fifth power of their mass. In contrast, when the dark matter mass is large, as is the case for the well-motivated weakly interacting massive particles (WIMPs), an exact or nearly exact symmetry is usually imposed in order to ensure its longevity. A notable exception is a spin 1/2 dark matter particle, quintuplet under $SU(2)_L$ and  with zero hypercharge, which can only decay into Standard Model particles via dimension-six operators~\cite{Cirelli:2005uq}.

In this paper we introduce a dark matter candidate which is naturally stable as a consequence of the Standard Model symmetries only. The dark matter particle is an antisymmetric tensor ${\cal{B}}_{\mu\nu}$, singlet under the Standard Model gauge group. Massive rank-two antisymmetric tensors are known to describe spin-1 modes~\cite{Ogievetsky:1967ij} and are actually not alien to particle physics: they arise naturally in interstring interactions (the so-called Kalb-Ramond fields~\cite{Cremmer:1973mg}) and have been used in QCD to describe the dynamics of spin-1 mesons, most notably the $\rho(770)$ and $a_1(1260)$, at low energies~\cite{Gasser:1983yg}. In fact, their specific features seem to provide a more natural description of mesons than regular vector fields. Here we will show that these specific features also play a key role in making them a good dark matter candidate.    

In full generality, a massive two-form field has six degrees of freedom, which under the Lorentz group project onto two distinct spin-1 field representations, mutually connected by a duality transformation~\cite{Cata:2014fna}. For concreteness we will choose the transverse representation, which corresponds to a massive Kalb-Ramond field, with well-defined ${\cal{C}}$ and ${\cal{P}}$ quantum numbers, specifically $J^{PC}=1^{+-}$. The Lagrangian of the model then reads:
\begin{align}
{\cal L}={\cal L}_{\rm SM}+{\cal L}_{\cal{B}}+{\cal L}_{\rm int}\;,
\label{eq:L}
\end{align}
where ${\cal  L}_{\rm SM}$ is the Standard Model Lagrangian and ${\cal L}_{\cal{B}}$ is given by
\begin{align}
{\cal{L}}_{\cal{B}}&=\frac{1}{4}\partial_{\lambda}{\cal{B}}^{\mu\nu}\partial^{\lambda}{\cal{B}}_{\mu\nu}-\frac{1}{2}\partial^{\mu}{\cal{B}}_{\mu\nu}\partial_{\rho}{\cal{B}}^{\rho\nu}\nonumber\\
&-\frac{m_{\cal{B}}^2}{4}{\cal{B}}_{\mu\nu}{\cal{B}}^{\mu\nu}-\lambda_{\cal{B}} {\cal{B}}_{\mu\nu}{\cal{B}}^{\nu\lambda}{\cal{B}}_{\lambda\rho}{\cal{B}}^{\rho\mu}\;,
\label{eq:L-B}
\end{align}
where the last term is only one representative of the quartic self-interaction terms. Since the cubic self-interaction ${\cal{B}}_{\mu\nu}{\cal{B}}^{\nu\lambda}{\cal{B}}_\lambda^{~\mu}$ identically vanishes for a two-form field ${\cal{B}}^{\mu\nu}$, this is the most general Lagrangian with operators up to mass-dimension four.

Finally, ${\cal L}_{\rm int}$ contains the interactions between ${\cal{B}}_{\mu\nu}$ and the Standard Model sector. Remarkably, the most general Lagrangian reduces to a single operator, namely
\begin{align}
{\cal L}_{\rm int}=c_{\cal{B}} {\cal{B}}_{\mu\nu}{\cal{B}}^{\mu\nu}(H^{\dagger}H)\;.
\label{eq:L-int}
\end{align}

All possible relevant interactions with Standard Model fermions and gauge bosons vanish in this model: for Dirac fermions, a potential operator ${\cal{B}}_{\mu\nu}{\bar{\psi}}_L\sigma^{\mu\nu}\psi_R$, with $\sigma^{\mu\nu}=\frac{i}{2}[\gamma^\mu,\gamma^\nu]$, is forbidden by the Standard Model symmetries. Regarding gauge bosons, a potential operator ${\cal{B}}_{\mu\nu} F_Y^{\mu\nu}$, where $ F_Y^{\mu\nu}$ is the field strength tensor associated with $U(1)_Y$, is a priori allowed by symmetries. However, (i) for an on-shell ${\cal{B}}_{\mu\nu}$ it vanishes and (ii) for an off-shell ${\cal{B}}_{\mu\nu}$ the tensor becomes non-dynamical, which leads to inconsistencies. In particular, the matching of the effective theory with a putative ultraviolet completion becomes impossible. Therefore, the Wilson coefficient of this operator must be set to zero (see~\cite{Ecker:2007us} for related examples). Finally, the interaction term ${\cal{B}}_{\mu\nu} {\cal{B}}^{\nu}_{~\lambda} F_Y^{\lambda\mu}$ identically vanishes due to ${\cal{B}}^{\mu\nu}$ being antisymmetric.

Absolute stability of ${\cal{B}}_{\mu\nu}$ of course depends on the form of specific ultraviolet completions. For instance, potential flavor interactions with right-handed neutrinos or vector-like fermions might induce ${\cal{B}}_{\mu\nu}$-decay (though not necessarily making it unstable on cosmological times, provided the new particles are sufficiently heavy). Nevertheless, if one restricts the interactions to Standard Model particles, a remarkable feature of our model is that its leading-order Lagrangian satisfies an accidental $Z_2$ symmetry, which automatically ensures the stability of ${\cal{B}}_{\mu\nu}$. Indeed, since any interaction with fermions and gauge bosons is forbidden, and the interactions with scalars are $Z_2$-preserving, one cannot generate $Z_2$-violating interactions. The argument applies to operators of arbitrary dimension at all orders in the perturbative expansion. As a result, potential higher-dimensional operators, such as ${\cal{B}}_{\mu\nu}{\bar{\psi}}_L\sigma^{\mu\nu}H\psi_R$, ${\cal{B}}_{\mu\nu}H^{\dagger}W^{\mu\nu}H$ or $\partial^{\mu}{\cal{B}}_{\mu\nu}\partial_{\rho}F_Y^{\rho\nu}$, come with vanishing Wilson coefficients and therefore pose no threat to stability. 

Furthermore, since the interaction of ${\cal{B}}_{\mu\nu}$ with the Standard Model proceeds through the Higgs portal, the particle has a WIMP character. In particular, its thermal population in the Early Universe is generated via scattering processes with Standard Model particles and a relic abundance is produced via thermal freeze-out. As a result, a singlet ${\cal{B}}_{\mu\nu}$ is absolutely stable and a viable dark matter candidate. The model is also remarkably simple, with very few parameters relevant for the dark matter phenomenology: $m_{\cal{B}}$ and the dimensionless parameter $c_{\cal{B}}$.

It is interesting to point out that the accidental $Z_2$ symmetry emerges only if ${\cal{B}}_{\mu\nu}$ is a singlet under the Standard Model gauge group. If ${\cal{B}}_{\mu\nu}$ were a doublet or a triplet under $SU(2)_L$, the $Z_2$-breaking operators ${\bar{\psi}}_L\sigma_{\mu\nu}{\cal{B}}^{\mu\nu}\psi_R$ and ${\mathrm{tr}}[{\cal{B}}_{\mu\nu}{\cal{B}}^{\nu\lambda}{\cal{B}}_\lambda^{~\mu}]$ should respectively be added to Eq.~(\ref{eq:L-int}). A doublet would decay into fermion pairs, but the triplet turns out to be stable to all orders: its decay into a Higgs pair via a ${\cal{B}}_{\mu\nu}$ loop is forbidden by $SU(2)_L$ invariance. In the following we will concentrate on the singlet case, but it is interesting to remark that the stability of the triplet is not linked to a $Z_2$ symmetry. In particular, this implies that the accidental $Z_2$ symmetry of Eqs.~(\ref{eq:L-B}) and (\ref{eq:L-int}) is in no way part of the Lorentz group. 

\begin{figure}[t]
  \centering
  \includegraphics[width=.75\columnwidth]{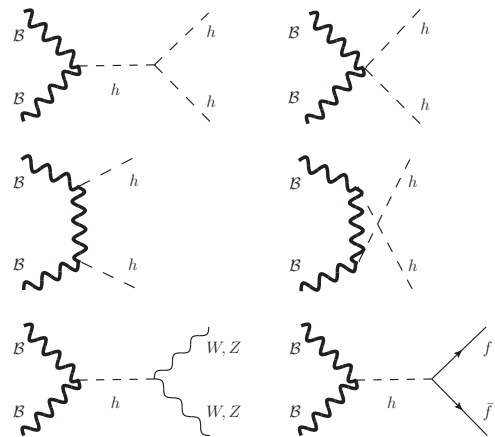}
 \caption{\small  Feynman diagrams contributing to the annihilation of antisymmetric tensor dark matter particles into Standard Model particles.}
\label{fig:annihilation-processes}
\end{figure}

The antisymmetric tensor relic abundance is determined by the annihilation rate into Standard Model particles. At tree level, the processes ${\cal B} {\cal B}\rightarrow f\bar f, WW, ZZ, hh$ are possible when kinematically accessible; the Feynman diagrams generating these processes are shown in Fig.~\ref{fig:annihilation-processes}. The annihilations into two weak gauge bosons or a fermion-antifermion pair are mediated by the s-channel exchange of a Higgs boson and resonantly enhanced when $m_{\cal B}\simeq m_h/2$. On the other hand, the annihilation into two Higgs boson is mediated by the s-channel exchange of a Higgs boson, the t- and u-channel exchange of an antisymmetric tensor and by the contact interaction ${\cal B}{\cal B} hh$.  The cross sections for the different channels can be readily calculated from the Feynman rules presented in the Appendix. The resulting expressions are however rather lengthy and will be presented in a separate publication. 

Due to the tensorial character of the field ${\cal{B}}_{\mu\nu}$ the total cross section has a pathological dependence with the center-of-mass energy $s$. Concretely, for large values of $s$ we find
\begin{align}
\sigma_{\rm tot}(s)&\simeq \frac{c_{\cal B}^2 s}{36\pi m_{\cal B}^4}
\left\{\frac{12 m_b^2}{s}\theta_b +\frac{12 m_t^2}{s}\theta_t +2\theta_W+\theta_Z \right. \nonumber \\
& \left. +\left[2+16\frac{c_{\cal B} v^2}{m_{\cal B}^2}+40
\left(\frac{c_{\cal B} v^2}{m_{\cal B}^2}\right)^2\right]\theta_h\right\}\;,
\label{eq:large-s}
\end{align}
where $\theta_X=\theta(\sqrt{s}-2m_X)$ is a Heaviside function accounting for the energy thresholds of the various particles and $v=246$ GeV is the Higgs field vacuum expectation value. As apparent from this equation, the cross section at large $s$ becomes constant for $m_{\cal{B}}<m_W$ and grows with $s$ when $m_{\cal{B}}>m_W$, thus requiring new physics at high energies to restore unitarity. This new physics could consist of new fermionic degrees of freedom charged under a confining gauge group, in analogy to the $\rho$ meson in QCD. We find numerically that the unitarity bound $\sigma<16\pi/s$ is not violated as long as $\sqrt{s}\lesssim 5 m_{\cal B}$, when $c_{\cal B}=0.1$. Hence, the new physics that unitarizes the theory is not expected to affect significantly the dark matter phenomenology, where the relevant processes take place at center-of-mass energies much smaller than the dark matter mass. 

We have calculated numerically the solution to the Boltzmann equation using the instantaneous freeze-out approximation \cite{Griest:1990kh}. Defining the variable $x_f\equiv m_{\cal B}/T_f$, the temperature $T_f$ at which the freeze-out takes place is approximately given by:
\begin{align}
x_f=\log\frac{0.038 c(c+2) g_{\rm int} M_{\rm Pl}m_{\cal B} \langle \sigma v\rangle}{g_*(x_f) ^{1/2} x_f^{1/2}}\;,
\label{eq:freeze-out-temperature}
\end{align}
where $c$ is a numerical factor, $c\approx 0.5$, $M_{\rm Pl}=1.22\times 10^{19}\,{\rm GeV}$ is the Planck mass, $g_{*}(x_{f})$ is the number of relativistic degrees of freedom at the freeze-out temperature and $g_{\rm int}$ is the number of dark matter internal degrees of freedom, in this case $g_{\rm int}=3$. We typically find $x_f=$20-30. Finally, the relic density of dark matter particles reads:
\begin{align}
	\Omega_{\cal B} h^{2} \simeq  \frac{1.07\times 10^{9}\,{\rm GeV}^{-1}}{J(x_{f})\, g_{*}(x_{f})^{1/2}\,M_{\rm Pl}}\,,
\label{eq:Omegah2}
\end{align}
where $J(x_{f})  =\int_{x_{f}}^{\infty}\,\langle\sigma  v\rangle x^{-2}\,\text{d}x\,$. Here $\langle\sigma  v\rangle$ is the thermally-averaged annihilation cross section, which can be readily calculated using the formalism presented in Ref.~\cite{Gondolo:1990dk}.

\begin{figure}[t]
  \centering
  \includegraphics[width=.75\columnwidth]{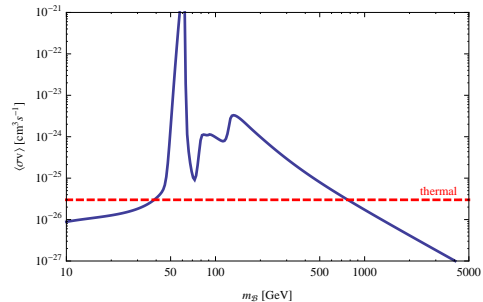}
  \caption{\small Thermally-averaged annihilation cross section as a function of the dark matter mass $m_{\cal B}$ at the temperature $T=m_{\cal B}/25$, assuming $c_{\cal B}=0.1$. We also show for reference the typical cross section at freeze-out required to reproduce the correct dark matter abundance,  $\langle \sigma v\rangle =3\times 10^{-26}\,{\rm cm}^3\,{\rm s}^{-1}$.}
\label{fig:sigmav}
\end{figure}

We show for illustration in Fig. \ref{fig:sigmav} the value of $\langle \sigma v\rangle$ as a function of the dark matter mass assuming  $c_{\cal B}=0.1$ and a value of the freeze-out temperature $T_f=m_{\cal B}/25$. The Higgs resonance is clearly visible at $m_{\cal B}\simeq m_h/2$, as well as the various thresholds where new channels contribute to the total annihilation cross section. It is important to note that, although the annihilation cross section either remains constant or grows with $s$, the thermally-averaged cross section remains finite, due to the Boltzmann suppression of the number of high energetic dark matter particles at the time of freeze-out.

\begin{figure}[t]
  \centering
  \includegraphics[width=0.75\columnwidth]{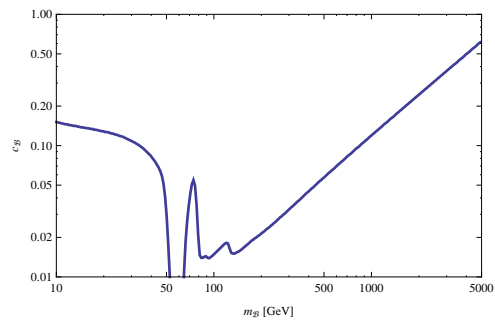}
  \caption{\small  Value of the coupling $c_{\cal B}$ as a function of the dark matter mass that reproduces the dark matter abundance $\Omega_{\cal B} h^2=0.1199 \pm 0.0027$ via thermal freeze-out.}
\label{fig:c-vs-mass}
\end{figure}

The dark matter relic abundance depends in this model on just two parameters: the dark matter mass and the Higgs portal coupling. It is then possible to determine the value of the coupling $c_{\cal B}$ as a function of the dark matter mass from the requirement that the observed dark matter abundance, $\Omega_{\cal B} h^2=0.1199 \pm 0.0027$~\cite{Ade:2013zuv}, is generated from thermal freeze-out. The value for $c_{\cal B}$ is shown in Fig. \ref{fig:c-vs-mass} and lies well within the perturbative regime in the range of masses of interest.

An approximate expression for the relic density can be derived in the limit $m_{\cal B}\gg m_h$. In this limit, resonance and threshold effects are negligible and the annihilation cross section at the time of freeze-out can be approximated by 
\begin{align}
\langle \sigma v\rangle \simeq \frac{5c_{\cal B}^2}{12\pi m_{\cal B}^2}\;.
\end{align}
Inserting this value in Eqs.~(\ref{eq:freeze-out-temperature}) and (\ref{eq:Omegah2}) and considering that $x_f\simeq$ 23-27 in the range $m_{\cal B}$=200 GeV-5 TeV ({\it c.f.} Eq.~\ref{eq:freeze-out-temperature}) we obtain
\begin{align}
c_{\cal B}\simeq 0.061\left(\frac{m_{\cal B}}{500\,{\rm GeV}}\right)\;.
\label{eq:cB-approx-largeM}
\end{align} 

\begin{figure}[t]
  \centering
  \includegraphics[width=0.75\columnwidth]{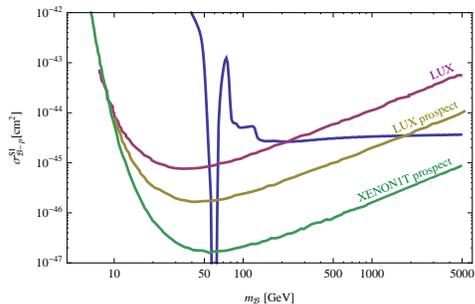}
  \caption{\small  Spin-independent scattering cross section of dark matter particles off protons, under the assumption that the dark matter population is generated via thermal freeze-out. We also show for comparison the present upper limit from the LUX experiment, as well as the projected sensitivity of the LUX and XENON1T experiments.}
\label{fig:sigmaSI}
\end{figure}

In the remainder of the paper we will turn our attention to the potential for direct detection of ${\cal{B}}_{\mu\nu}$. Antisymmetric tensor dark matter particles interact with the partons inside the nucleus via the Higgs portal, thus opening the possibility of detecting these particles in direct search experiments. The spin-independent interaction cross section can be approximated by~\cite{Barbieri:1988zs}
\begin{align}
\sigma^{\rm SI}_{{\cal B}-p}\simeq \frac{c^2_{\cal B} f_N^2}{4\pi} \frac{\mu^2 m_p^2}{m_h^4 m_{\cal B}^2}\;,
\label{eq:SI}
\end{align}
where $f_N \simeq  0.3$ is the nucleonic matrix element and $\mu=m_p m_{\cal B}/(m_p+m_{\cal B})$, $m_p$ being the proton mass. The resulting cross section is shown in Fig.~\ref{fig:sigmaSI} for the value of $c_{\cal B}$ that leads, for a given dark matter mass, to the observed dark matter abundance via thermal freeze-out (see Fig.~\ref{fig:c-vs-mass}). For $m_{\cal B}\gg m_h$, and using Eqs.~(\ref{eq:cB-approx-largeM}) and (\ref{eq:SI}), we find $\sigma^{\rm SI}_{{\cal B}-p}\simeq 3.5\times 10^{-45}\,{\rm cm}^2$.  We also show for comparison the present limit from the first analysis of the LUX experiment based on 85.3 live-days of data~\cite{Akerib:2013tjd}, as well as the projected sensitivities of the LUX~\cite{Akerib:2012ys} and XENON1T~\cite{Aprile:2012zx} experiments. As apparent from the plot, the present LUX limit allows the region  $m_{\cal B}\lesssim 230$ GeV, as well as the region of the Higgs resonance $m_{\cal B}\simeq 60$ GeV. Furthermore, the LUX (XENON1T) experiment will probe dark matter masses below 2 TeV (20 TeV) in this model.

To summarize, in this paper we have introduced a new dark matter spin-1 candidate, realized as an antisymmetric rank-two tensor ${\cal{B}}_{\mu\nu}$, which transforms as a singlet under the Standard Model symmetries. Remarkably, these symmetries alone induce an accidental discrete $Z_2$ symmetry of the Lagrangian that ensures the stability of ${\cal{B}}_{\mu\nu}$. Moreover, the interactions with Standard Model fields occur via the Higgs portal, thereby generating a relic population through thermal freeze-out. As a result, the singlet antisymmetric tensor is a perfect candidate for WIMP dark matter. 

We have also investigated the signatures of this dark matter candidate in direct detection experiments and we have found that the present LUX limit on the spin-independent interaction cross section restricts the dark matter mass to be larger than 230 GeV, or else to lie at $\sim 60$ GeV. The projected XENON1T experiment will extend the search for this particle up to a mass of 20 TeV.

\vspace{0.5 cm}
We would like to thank Lei Feng, Thomas Hambye,  Jernej Kamenik, Geraldine Servant, Shahin Sheikh-Jabbari and Jure Zupan for useful discussions. This work was partially supported by the DFG cluster of excellence ``Origin and Structure of the Universe'' and by the ERC Advanced Grant project ``FLAVOUR''(267104).  \\

\subsection*{Appendix}
The propagator of ${\cal{B}}_{\mu\nu}$ follows from the inversion of the kinetic term in Eq.~(\ref{eq:L-B}) and can be expressed as

\begin{equation}
\Delta_{\mu\nu;\lambda\rho}^{\cal{B}}(q)=\frac{2}{q^2-m_{\cal B}^2}\left[I_{\mu\nu;\lambda\rho}-\frac{q^2}{m_{\cal B}^2}P^L_{\mu\nu;\lambda\rho}\right]\;.
\label{eq:propagator}
\end{equation}
This is the natural extension of the Proca propagator to rank-two tensors, where
\begin{align}
P^L_{\mu\nu;\lambda\rho}&=\frac{q_{\mu}q_{\lambda}}{2q^2}g_{\nu\rho}-\frac{q_{\mu}q_{\rho}}{2q^2}g_{\nu\lambda}-\frac{q_{\nu}q_{\lambda}}{2q^2}g_{\mu\rho}+\frac{q_{\nu}q_{\rho}}{2q^2}g_{\mu\lambda}\;,\nonumber\\
I_{\mu\nu;\lambda\rho}&=\frac{1}{2}(g_{\mu\lambda}g_{\nu\rho}-g_{\mu\rho}g_{\nu\lambda})\;.
\end{align}
In terms of single particle creation, the propagator of Eq.~(\ref{eq:propagator}) corresponds to the normalization
\begin{align}
\langle 0| {\cal{B}}_{\mu\nu}|b(q,\lambda)\rangle&=\frac{i}{m_{\cal{B}}}\epsilon_{\mu\nu\alpha\beta}\,\varepsilon^{\alpha}_{(\lambda)}q^{\beta}\;.
\end{align}
The Higgs portal interaction of Eq.~(\ref{eq:L-int}) corresponds to the following Feynman rules:
\begin{figure}[h!]
  \centering
  \includegraphics[width=0.65\columnwidth]{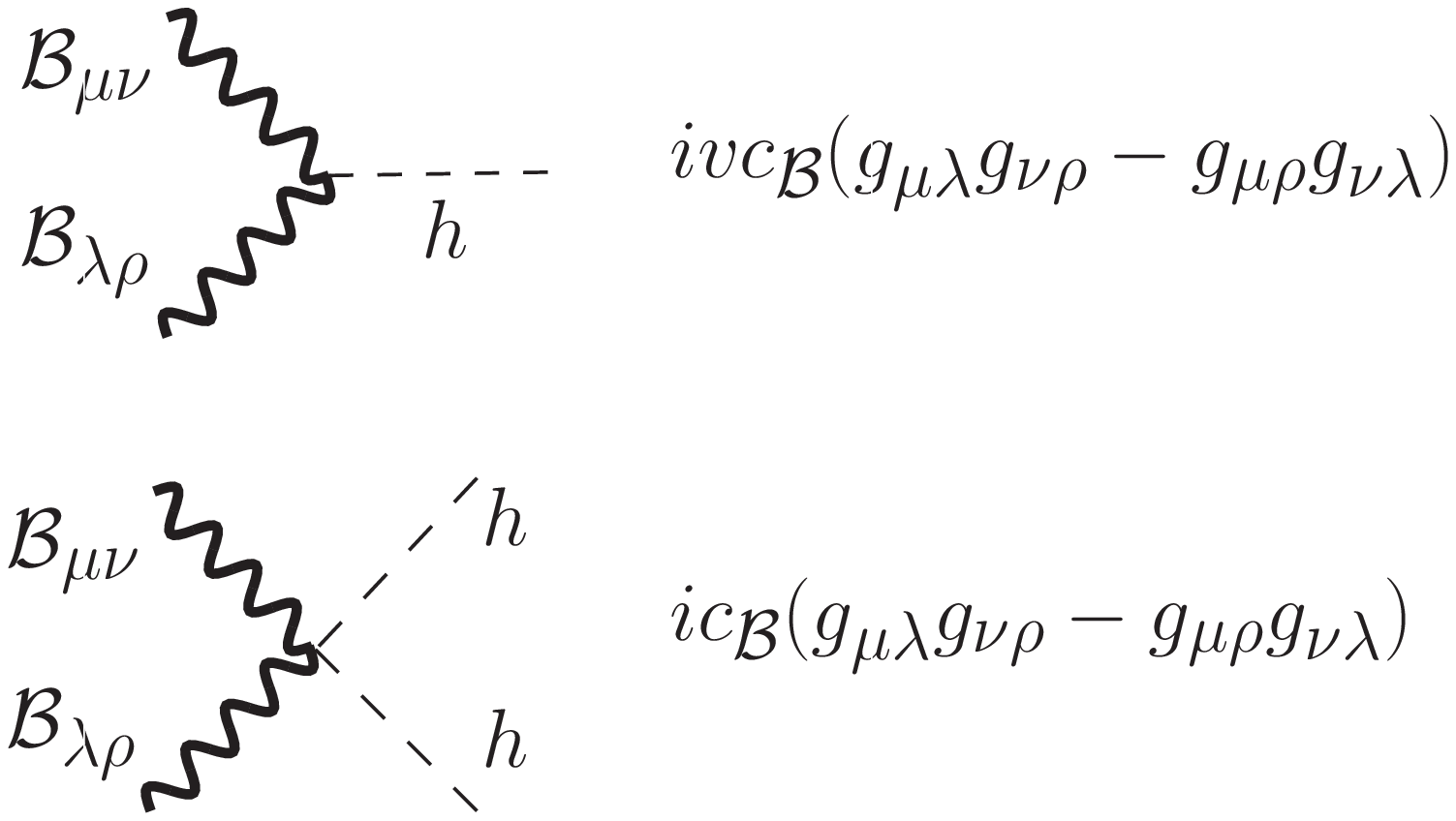}
\label{fig:Feynman-rules}
\end{figure}
\vskip -0.8cm


\end{document}